\documentclass[traditabstract]{aa}
\usepackage{graphicx}
\usepackage{txfonts}
\usepackage{natbib} 
\usepackage[urlcolor=blue,colorlinks=true,linkcolor=blue,citecolor=blue]{hyperref}
\begin{document}

\title{Astrometric search for a planet around VB 10\thanks{Based on observations made with ESO telescopes at the La Silla Paranal Observatory under programme ID 283.C-5024}}

\author{P. F. Lazorenko \inst{1} 
  \and J. Sahlmann \inst{2}
  \and  D. S\'egransan \inst{2}
  \and P. Figueira  \inst{2,5}
  \and  C. Lovis \inst{2}
  \and  E. Martin \inst{3} 
  \and  M. Mayor \inst{2}
  \and  F. Pepe \inst{2}
  \and  D. Queloz \inst{2}  
  \and F. Rodler \inst{4}
  \and  N. Santos \inst{5,6}
    \and  S. Udry \inst{2}
   	}

     \institute{Main Astronomical Observatory,
          National Academy of Sciences of the Ukraine,
             Zabolotnogo 27, 03680 Kyiv, Ukraine
     \and  Observatoire de  Gen\`eve,  Universit\'e de Gen\`eve, 
         51 Chemin des Maillettes, 1290 Sauverny, Switzerland   
     \and  INTA-CSIC Centro de Astrobiolog\'{\i}a, 28850 Torrej\'on de Ardoz, 
          Madrid, Spain
     \and Instituto de Astrof\'{\i}sica de Canarias, C/ V\'{\i}a L\'actea s/n, 
          E-38200 La Laguna (Tenerife), Spain
     \and Centro de Astrof\'{i}sica, Universidade do Porto, Rua das Estrelas,
              4150-762 Porto, Portugal
    \and Departamento de F\'isica e Astronomia, Faculdade de Ci\^encias, 
Universidade do Porto}

%   \date{Received September 15, 1996; accepted March 16, 1997}
   \date{Received ; accepted  }                                                            

   \abstract
{We observed VB~10 in August and September 2009 using the FORS2 camera of the VLT with the aim of measuring its  astrometric motion and of probing the presence of the announced planet VB 10b. We used the published STEPS astrometric positions of VB~10 over a time-span of 9 years, which allowed us to compare the expected motion of VB~10 due to parallax and proper motion with the observed motion and to compute precise deviations. The achieved single-epoch precisions of our observations are about 0.1~mas and the data showed no significant residual trend, while the presence of the planet should have induced an apparent proper motion larger than 10 mas yr$^{-1}$. Subtraction of the predicted orbital motion from the observed data produces a large trend in position residuals of VB~10. We estimated the probability that this trend is caused by random noise. Taking all the uncertainties into account and using Monte-Carlo resampling of the data, we are able to reject the existence of VB~10b with the announced mass of 6.4~M$_{\rm J}$ with the false alarm probability of only $5 \cdot 10^{-4}$. A 3.2 M$_{\rm J}$ planet is also rejected with a false alarm probability of 0.023. }

\keywords{astrometry  -- technique: high angular resolution --  planetary systems -- stars: individual: VB 10}
              
     \maketitle
\section{Introduction}
The detection of a planetary-mass companion to the nearby M8 ultracool dwarf \object{VB~10} (\object{GJ~752B}) was announced by \cite{Pravdo}. Using ground-based astrometric observations with the STEPS camera of the 5 m Palomar telescope  {obtained over nine years}, the authors derived a full astrometric solution of the system that suggested the presence of a planetary companion with a 271~ {day} orbital period and a 6.4 $M_{\rm J}$ mass. Even though the \object{VB~10b} astrometric signature is large with a peak-to-peak value of 8~mas, the detection is debatable due to imperfect orbital sampling and comparably low astrometric accuracy of   {the} individual measurements.

\cite{Zapatero} presented high-resolution near-infrared observations of VB~10 obtained between 2001 and 2008 with the NIRSPEC instrument on the Keck II telescope. They emphasized on the necessity to have a better sampling of the orbital phase to precisely constrain the orbital parameters and the individual masses of the  {system}. \cite{Bean} used the near-infrared CRIRES spectrograph and did not detect the planet's signature in radial-velocity  {(RV)}. \cite{Anglada} performed a joint analysis of precise RV observations over 175 days with MIKE/Magellan and ESPaDOnS/CFHT spectrographs, RV measurements by \cite{Zapatero}, and astrometric data by \cite{Pravdo}.  {They showed} that the observed astrometric motion is not due to an unseen companion.

We present results of the astrometric search for the planet using the FORS2 camera of the VLT, the only instrument which can achieve astrometric precision of individual measurements of 0.05--0.1 mas \citep{Lazorenko2009}.  {Despite the short observation period of two months, }we obtained data sufficiently good to verify the hypothesis on the possible existence of VB~10b.

\section{Observations}
Observations  were made with the FORS2/VLT camera \citep{Appenzeller} in imaging mode during five nights from 2 August to 25 September 2010.  {The images cover a $4.2 \times 4.2$\arcmin field of view with a pixel  {(px)} scale of $\sim$0.1\arcsec/px. For each night, we obtained 21 to 72 frames of 7 s exposures with the $R_{\rm special}$ filter.} Seeing varied from 0.55\arcsec to 0.9\arcsec. To reduce differential chromatic refraction (DCR) of the atmosphere, observations were made near meridian.

 {Because the observed motion of VB~10  in the sky is large (over 2 px), we had to know the precise pixel scale to convert pixel to arcsec units without loss of accuracy.}  {From} CMC14 and NOMAD catalogue positions of stars in our reference frame, we derived acceptably precise scales of $0.12538 \pm 0.00007$~\arcsec/px and $0.12567 \pm 0.00005$~\arcsec/px in RA and Decl, respectively.

\section{Data reduction and analysis}
{ Our null hypothesis was that the planet VB~10b exists. To test this hypothesis, we computed the residuals of our measurements compared to the position of VB~10 expected from the combination of proper, parallactic, and orbital motions. By determining the probability that the measured residuals are compatible with the expected mean value of zero, we could confirm or reject the null hypothesis. We also performed the corresponding computations assuming that the planet does not exist. 

In all cases, we considered that the astrometric signal $\Psi$ in VB~10 position induced by the planet} is defined by the orbital elements of VB~10b given in the discovery paper by \cite{Pravdo}. The predicted orbital motion over the measurement timespans of 17 and 54 days is  0.48 and 1.95 mas in RA and 0.92 and 3.28 mas in Decl, respectively. Compared to the astrometric precision of FORS2 of about 0.1~mas per epoch, these displacements are large and should be detected in this study.

\subsection{Photocentre determination}
Raw  {images} were flat-fielded and bias-subtracted to exclude pixel-to-pixel variations of the CCD sensitivity. To increase the number of reference stars, we measured all star images with $R= 14$--21. At this faint end, the star field is very crowded. This forced us to improve our procedure of photocentre computation, which initially was developed for isolated stellar images \citep{Lazorenko2006}. We scanned images obtained under various seeing conditions  {and made} a detailed census  {of} star positions at subpixel precision,  {of} fluxes, and  {of} the point-spread function variation across the CCD. This information was used to accurately model and subsequently subtract the contamination of background counts caused by distant stars. Wings of star profiles at distances up to 50~px were approximated by exponential function with free parameters smoothly varying over the CCD.  {The} photocentres $\hat x$  {and} $\hat y$ were computed  {by} fitting star profiles in a $11\times11$~px window with a model with 12 free parameters  {and} an auxiliary oscillating function in the central $5\times5$~px window \citep{Lazorenko2006}. Computations based on the Levenberg-Marquardt numerical algorithm of the least squares fit were found to give sufficiently good results for stars of approximately equal brightnesses  {at} separations  {larger than} about 10 px.

\subsection{Astrometric model}
Reductions were based on the method  {previously} applied to FORS1/2 observations \citep{Lazorenko2007, Lazorenko2009}.
The method was shown to efficiently mitigate atmospheric image motion, geometric field distortion, DCR, and other effects, thus ensuring precision stable over time scales of a few days to a few years.  {For every star in the field, i.e. VB~10 and the reference stars,} the measured  {photocenter} positions $\hat x_m$ and $\hat y_m$  {in} frame $m$
 {at} time $t$ were represented by the model
\begin{equation}
\begin{array}{l}
\label{eq:model}
     x_0+ \Phi_{km}^{\{ x \}}(x,y) + \mu_x t+ \pi p_x +
       \rho \tan z  \sin \gamma+  d \tan z_{\rm L} \sin \gamma   \\
\qquad{}  \qquad{}        = \hat x_m  - \Psi(t).
\end{array}
\end{equation}
The left side contains the free model parameters and the model function $\Phi_{km}^{\{ x \}}(x,y)$. The expression for $y$-data is similar but contains $\cos  \gamma$ instead of $\sin \gamma $. Here, $x_0$ is a zero point, $k$ is the mode (arbitrary even integer, usually from 4 to 16), $\Phi_{km}^{\{ x \}}$ is a polynomial in $x$ and $y$ of order $k/2-1$ that models  {the} sum of atmospheric image motion and geometric distortion for each frame $m$.  {The parameter} $\mu_x$ is the proper motion, $\pi$ is the parallax, and $p_x$ is the parallax factor in $x$.  {The} displacement of the star image due to DCR is modelled by a term with leading parameter $\rho$, which depends on the star's colour,  {and containing the zenith distance $z$ and the angle $\gamma$ between a direction to zenith and $y$-axis.} The next term describes an image displacement  {opposite} to that of DCR and introduced by the longitudinal atmospheric dispersion compensator (LADC) of the VLT \citep{Avila}. This displacement also depends on the star colour via  {the parameter} $d \approx - \rho$. Both $\rho$ and $d$ are free model parameters  {and} their computation does not require external colour data. The LADC is automatically adjusted to the average zenith distance $z_{\rm L}$ over a given series of frames by setting a distance $b$ between its two prisms to $b \sim \tan z_{\rm L}$.  {Finally,} $\Psi(t)$ represents the  {induced} orbital motion { of VB~10 if the planet exists. For reference stars, $\Psi(t)=0$.}

Equation~\ref{eq:model} defines a system of equations in the combined \{$x$, $y$, $t$\}-domain for which the unique solution is derived under the condition that the model parameters of reference stars are orthogonal to each basic function of $\Phi_{k}$.  {It} is solved iteratively for all reference stars available in a circular region of radius $R_k$ which increases with $k$ and is centered on the target VB~10. The optimal field size $R_k$ is the size at which the noise from a reference field is equal to the noise from atmospheric image motion. The number of reference stars depending on $k$ varies from a few dozens to 500. Due to the large number of reference stars, the function $\Phi$ accurately reproduces the coordinate grid distortion introduced by image motion. Solutions for $\Phi_{km}$ at each mode $k$ are then used to form equations for the target in the time domain only. The solution of this new set of equations yields the target's model parameters and position deviations from the model. The final output is the average obtained from the solutions at all modes $k$.

\begin{figure}[t ]
\begin{tabular}{@{}c@{}c@{}c@{}}
{\includegraphics*[bb = 61 57 172 169, width=3cm,height=3cm]{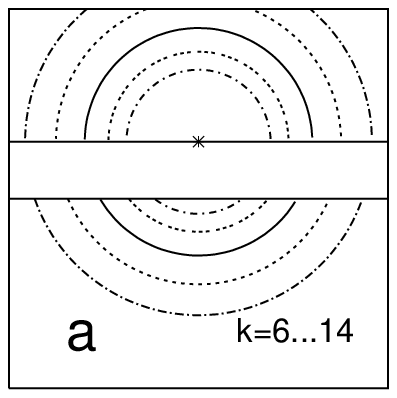}} &
{\includegraphics*[bb = 61 57 172 169, width=3cm,height=3cm]{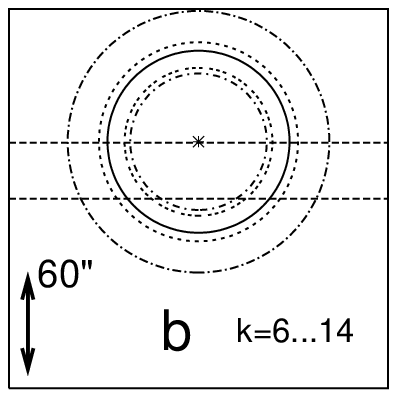}} &
{\includegraphics*[bb = 61 57 172 169, width=3cm,height=3cm]{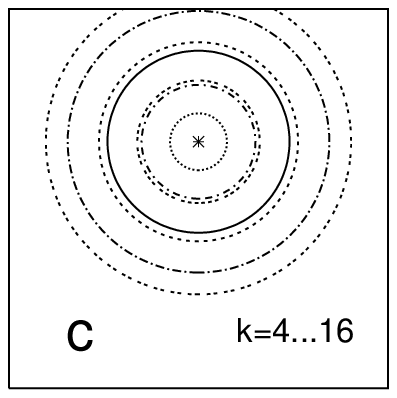}}
\end{tabular}
\caption {Circular reference fields
increasing in size with increasing modal number $k$
in the field of view of FORS2 in the case of:
{\bf a)} the full data set when an area
unavailable on 2 Aug frames was cut away at all nights;
{\bf b)} the same but the problem area cut only at 2 Aug frames;
{\bf c)} the short data set not using 2 Aug frames.
         }
\label{fov}
\end{figure}
%%%%%%%%%%%%%%%%%%%%%%%%%%%%%%

\subsection{Comparison of STEPS and FORS2 reference frames}\label{sec:refframes}
{Because of different zero-points,} STEPS proper motion cannot be {directly} applied to FORS2 astrometry. The STEPS reference frame is given by 15 bright stars. For the FORS2 reference frame, we use both these (except STEPS star Nr. 10 which is a binary) and a number of much fainter stars. The system of proper motions is defined under the condition $\sum \mu_i w_i = 0$ where summation is taken over all reference stars $i$ with weights { $w_i$ approximately proportional to their brightness} \citep{Lazorenko2009}. In particular, this condition is valid for the proper motions of the STEPS stars which, due to their high brigthness (they contribute to over 40\% of the total light flux), are the basis of the FORS2 reference frame. Hence, one may expect that $\sum \mu_i w_i/\sum w_i$ (the weighted mean proper motion in the system of FORS2 proper motions) taken over these stars, is nearly zero. The computed value of this sum (which is the difference of proper motion zero-points) was found to be small: $-0.1 \pm 1.0$ and $+0.3 \pm 1.0$ mas yr$^{-1}$ in RA and Decl. Due to the low precision, we did not apply it but included it in the error budget assuming that the uncertainty in the proper motion zero point is $\sim$1~mas~yr$^{-1}$. Thus, both proper motion systems are well consistent, however at time scales of $\sim 0.1$~yr only.

\subsection{Short and full dataset}
{The data reduction of the first epoch (2 Aug) was problematic because of incorrect telescope pointing,}
owing to which all stars in a 300~px wide area just below VB~10 (Fig.~\ref{fov}a) were imaged to another CCD chip and could not serve as reference at this particular night. We dealt with this problem in two ways. In the first version, we put aside all stars within this area from frames of all nights, using only stars outside of this area as reference objects. {The sizes and shapes of the reference fields in this case are} shown in Fig.~\ref{fov}a by five circular segments, each corresponding to $k$ increasing from six (small {radius}) to 14 (large {radius}). Alternatively, we used circular reference areas, which for the 2 Aug night were of the same $R_k$ size but with {the} lower half vignetted (Fig.~\ref{fov}b). We also examined a short dataset without {the} 2 Aug epoch with circular reference area (Fig.~\ref{fov}c). The {asymmetry} of the first two cases degraded the {obtained} precision in comparison to the symmetric configurations of the short dataset.

We have obtained a solution for all three cases presented in Fig.~\ref{fov}. For the final result, we present two solutions. One is obtained from the short dataset (case 'c', {MJD between 55082 and 55099 days}), and the second one is obtained as the average of the results in case 'a' and 'b'. The latter solution is referred to as obtained from the full dataset {(MJD between 55045 and 55099 days)}.

\subsection{Recovering LADC positions}
{VB~10 is very red and} differs much in colour from the reference stars. Therefore, the DCR displacement of this star is very large {and} about 30 mas in Decl and 10 mas in RA. This is taken into account by the free model parameters $\rho$ and $d$, {but their adjustment} requires knowledge of the LADC separation $b$.

\begin{figure}[t]
%\resizebox{\hsize}{!}{\includegraphics*[bb = 53 49 252 280, height= 3.5cm]{b1_best.eps}}
%{\includegraphics*[bb = 53 49 252 266, width=\linewidth, height= 9.5cm]{b1_best.eps}}
{\includegraphics*[bb = 53 49 252 266, width=\linewidth]{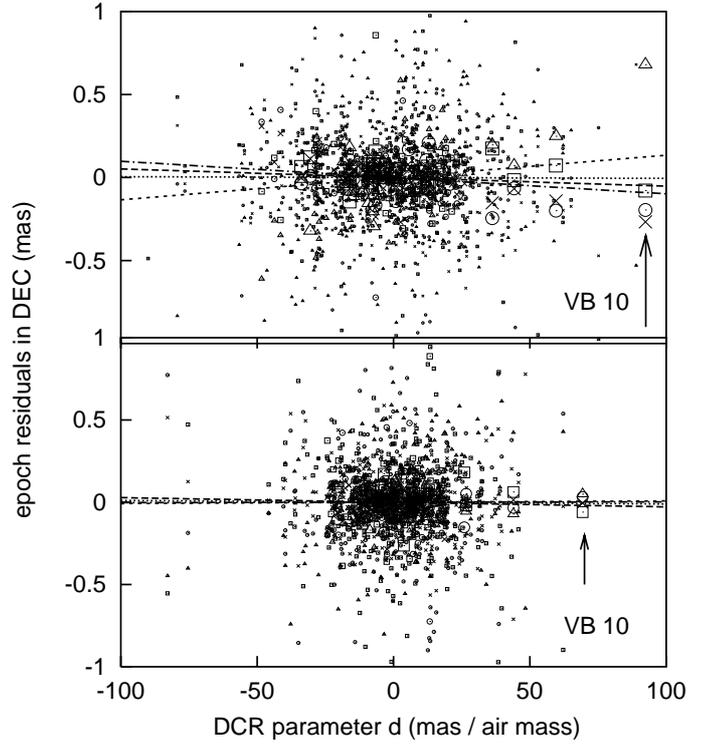}}
\caption{Distribution of epoch position residuals $\langle y \rangle $ of field stars (dots of different type) as a function of $d$ for four September nights and linear dependences of $\langle y \rangle$ on $d$ (lines of different type) for each night. {\it Upper panel}: with initial LADC  separations $b$. {\it Lower panel}: the same with
recovered $b$ values. Size of dots is proportional to the star brightness.}
\label{b_ladc}
\end{figure}
%%%%%%%%%%%%%%%%%%%%%%%%%%%%%%

Information on $b$ {was} not part of the obtained fits headers and was not accessible to us. Therefore we had to solve the inverse problem of recovering LADC separations from the observations. Because $b$ is fixed for a series of frames of a single night, we had to {determine a small number of values for b (one value per night).} Our approach is based on the observation that incorrect values of $b$ bias the average night position residuals $\langle x \rangle$ and $\langle y \rangle$ of field stars. According to Eq.~\ref{eq:model}, this bias {linearly depends on the star colour parameter $d$ during one given night}. This is illustrated by Fig.~\ref{b_ladc} which shows the distribution and linear dependencies of  $\langle y \rangle$  on $d$ for field stars in the case of the short dataset (the effect is largest in $y$).

Because all observations were made at small hour angles within $\pm$0.7 h, we initially assumed that LADC was always set to the separation corresponding to the meridian. By applying small corrections to these initial values of $b$, we iteratively reached a solution without dependence of $\langle y \rangle$ on colours (lower panel of Fig.~\ref{b_ladc}). {For these computations, we used  all stars within the entire field of view of  FORS2, processing them as {\it target} objects relative to their own subsets of reference stars.} Being based on {\it field} stars only, this procedure produced a much smaller dispersion of $\langle y \rangle$ values for VB~10 and for reference stars with large colours.  Corrections to $b$ were small and within $\pm 4 $\% of the initial values and were found with precision corresponding to $\pm 0.025$ mas error in $\langle y \rangle$ for VB~10.  For the full dataset, precision degraded  to $\pm 0.068$ mas because { we could not use stars in the problematic 300~px area below VB~10 (Fig.~\ref{fov}a).}

\subsection{Treatment of colour-dependent terms}\label{sec:col}
An essential drawback of the above procedure is that the recovery of $b$ {introduces} small colour-dependent terms $\mu'(d) $ and $\pi' (d)$, which are similar to proper motion and parallax. {This is because the above iterations do not garantee convergence to the actual values of $b \sim \tan z_{\rm L} $. For example, the restored value of $\tan z_{\rm L} $ may differ from its actual value by a term which progressively changes in time as $\tilde{\mu} t \tan^{-1} z_{\rm L} {\langle \sin \gamma \rangle}^{-1} $, where $\tilde{\mu}$ is an arbitrary constant and $\langle \sin \gamma \rangle $ is the average value of $ \sin \gamma $ during the given night. Hence the term $  d \tan z_{\rm L} \sin \gamma$  in Eq.~\ref{eq:model} generates the term $ d \tilde{\mu} t \sin \gamma {\langle \sin \gamma \rangle}^{-1} \!\approx \!d \tilde{\mu} t $  linearly dependent on $d$ and which therefore can be treated as an extra image motion $\mu'(d)=  \tilde{\mu} d$. The term $\mu'(d)$ compensates for the linear change of  $\tan z_{\rm L}$ in time. Similarly, we may assume that the restored values of $\tan z_{\rm L} $ contain terms proportional to $p_x$ and $p_y$. In this case, the solution of Eq.~\ref{eq:model} for parallax should contain the compensating colour-dependent term $\pi'(d)$.  }

When recovering $b$ values, we cannot control the amplitude of $ \tilde{\mu}$ and {of the equivalent parameter $ \tilde{\pi}$  related to parallax. However, they can be detected as proper motion and parallax dependence on colours (i.e. on $d$) which we model as a linear trend in proper motions of field stars and statistically correct for it}.

{The} strong correlation between $t$, $p_x$, and $p_y$ does not allow us to determine {separately $\mu'(d)$ and $\pi'(d)$}. However, this is not required because {for short times $p_x \! \sim \! t$, we can approximate  the sum of the proper motion and parallax displacement $\mu'_x t+\pi' p_x$  by $\mu'_x t$, where  $\mu' $ is the new effective quantity that substitutes for both $\mu'$ and $\pi'$. Thus processing the short dataset, we find and use only a single colour term $\mu'(d)$. The treatment is similar for the $y$ components.} Subtraction of $\mu' t$ from the measured positions thus {simultaneously eliminates the colour term} $\pi'$ {(see next section)}. The value of this correction for VB~10 in Decl is $\mu'=29 \pm 4.5$ mas yr$^{-1}$ for the full and  $\mu'=27 \pm 2.9$ mas yr$^{-1}$  for the short dataset, respectively. In RA, the corrections are an order of magnitude smaller.

For the full dataset, the above approximation of $\mu'_x t+\pi' p_x$  by $\mu'_x t$ is {still} valid for reference stars, { most of which have moderate $d$ and therefore very small colour-induced parallaxes $\pi'$}. But it is not sufficiently precise for VB~10, because of its large colour term $d$. Consequently, $\pi'$ is not {accurately eliminated} by applying the correction  $\mu'(d)$ as in the case of the short dataset. Therefore, we had to treat the term $\pi'$ of VB~10 as a free model parameter.

\begin{figure}
\begin{center}
\includegraphics[width=\linewidth]{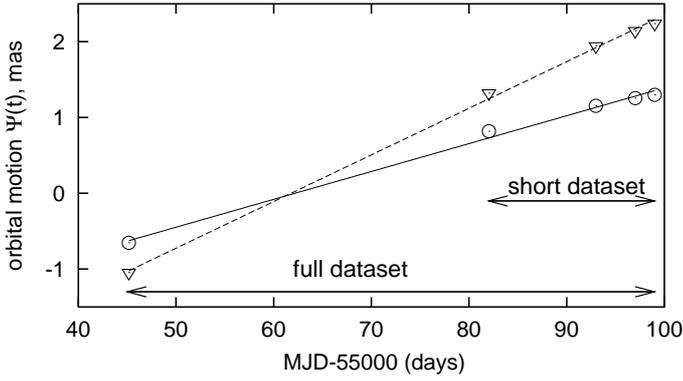}
\caption{Orbital motion $\Psi(t)$ of VB~10 in RA (circles) and Decl (triangles) {for} the full and short FORS2 datasets and the linear approximations {(solid and dashed lines)} of these motions.}
\label{psi}
\end{center}
\end{figure}

\subsection{Subtraction of parallax and proper motions}\label{sec:ppm}
Since the FORS2 observations {cover a small fraction ($<\!20$\,\%) of} the orbital period, $\Psi(t)$ can be {approximated as }a linear function of $t$:
\begin{equation}
\label{eq:appr}
\Psi_{\alpha}(t) \sim   Q_{\alpha} t,
\qquad{}  \qquad{}
\Psi_{\delta}(t) \sim    Q_{\delta} t,
\end{equation}
where $Q_{\alpha}= { \partial \Psi_{\alpha}(t) }/{ \partial t  }$ and $Q_{\delta}=  { \partial \Psi_{\delta}(t) }/{ \partial t  } $.
With the orbital elements given by \cite{Pravdo}, {we obtained} $Q_{\alpha}{\rm (PS)}=13.4$ and $Q_{\delta}{\rm (PS)}=22.4$ mas yr$^{-1}$ (Fig.~\ref{psi}). Astrometric acceleration {terms} (deviations from the linearity) {are smaller than $0.1$ mas for the full dataset, thus are negligible.} For the short dataset, they are even smaller than 0.02~mas.

{In addition}, $\Psi(t)$ is approximately a linear function of the parallax factors $p_x(t)$ and $p_y(t)$, which themselves {have an approximately linear time-dependence.} This causes a strong correlation between parallax, proper motion, and orbital motion, thus makes Eq.~\ref{eq:model} degenerated. Therefore, we subtracted parallax and proper motion from the measured positions of VB~10. Precise values of these parameters were found based on the published STEPS astrometric measurements which cover a 9 years long period. The best fit of STEPS data yielded $\mu_{\alpha} \cos (\delta) = -586.8 \pm 0.2$ mas yr$^{-1}$, $\mu_{\delta}= -1361.0 \pm 0.2$ mas yr$^{-1}$, and $\pi=168.0 \pm 1.2$ mas.{ These values} are very close to the estimates given by \cite{Pravdo} and \cite{Anglada}.

\begin{figure}[htb]
\begin{tabular}{@{}c@{}}
{\includegraphics*[bb = 53 59 336 181, width=\linewidth, height=115pt]{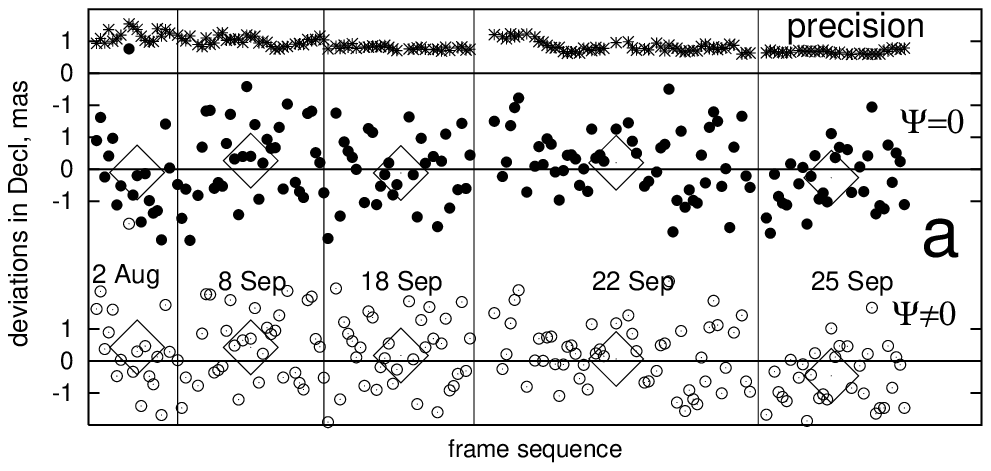}}  \\
{\includegraphics*[bb = 53 59 336 193, width=\linewidth, height=125pt]{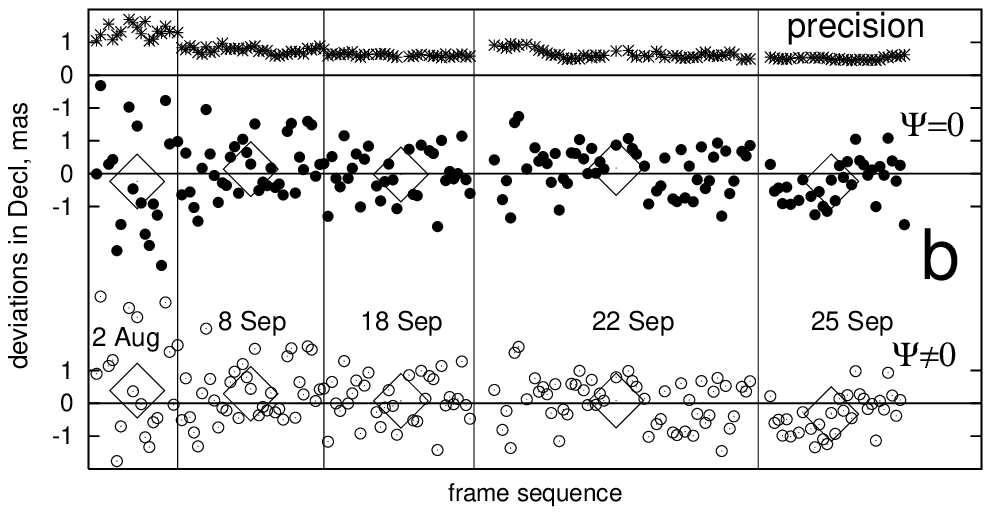}} \\
{\includegraphics*[bb = 53 50 336 168, width=\linewidth, height=115pt]{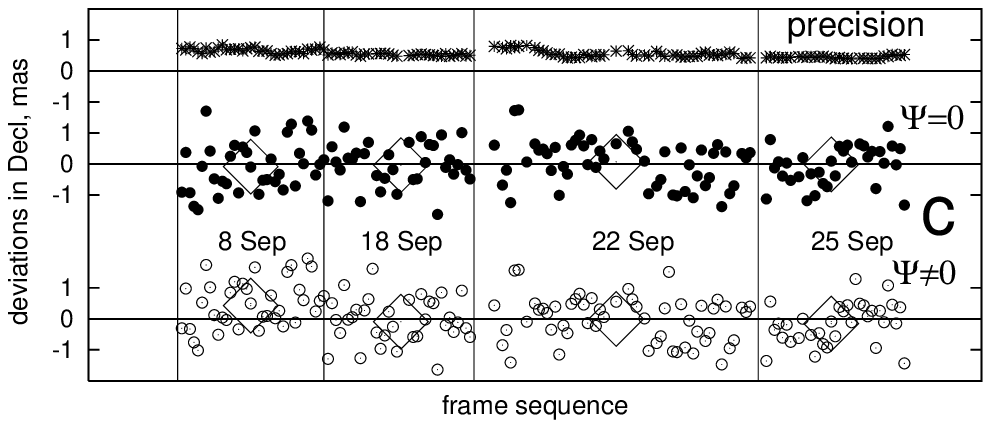}} \\
\end{tabular}
\caption{Frame-to-frame deviations in Decl of VB~10 from its model motion at zero ($\Psi=0$, filled circles) and non-zero planetary signal ($\Psi \neq 0$, open circles), the epoch average deviations (large diamonds), and the single-frame precisions $\sigma_{\rm 1}$ (asterisks) for the cases {\bf 'a', 'b', and 'c'} of reference field configurations shown in Fig.\ref{fov}.} 
\label{abcplot}
\end{figure}
              
\subsection{Position residuals of models with and without planet}\label{sec:posres}
{Subtracting the contributions} of parallax and proper motion from {the measured} positions of VB~10 resulted in great simplification of Eq.~\ref{eq:model}, which for VB~10 took the form
\begin{equation}
\begin{array}{l}
\label{eq:model2}
   x_0   +   \rho \tan z  \sin \gamma+    d \tan z_{\rm L} \sin \gamma 
         + \{ \pi' p_x \} \\
\qquad{} = \hat x_m - \mu_x t - \pi p_x - \mu'_x t -\Phi_{km}^{\{ x \}} (x,y) - \Psi_x(t)  \\
   y_0   +   \rho \tan z  \cos \gamma+    d \tan z_{\rm L} \cos \gamma
         + \{ \pi' p_y \}  \\
\qquad{} = \hat y_m - \mu_y t - \pi p_y - \mu'_y t -\Phi_{km}^{\{ y \}}(x,y)  - \Psi_y(t)  \\
\end{array}   %\qquad{}
\end{equation}
with only four free parameters $x_0$, $y_0$, $\rho$, and $d$ in the case of the short dataset. For the full dataset, it incorporates the extra parameter $\pi'$ (the term in curly braces, see Sect.~\ref{sec:col}). The right side of Eq.~\ref{eq:model2} contains the measured coordinates $\hat x_m$ and $\hat y_m$, the in-plate solutions $\Phi_{km}^{\{ x \}}(x,y)$ and $\Phi_{km}^{\{ y \}}(x,y)$ derived from the reference stars, and all other corrections. 

{ There is a correlation between the orbital signal and the night average angles $\langle \sin \gamma \rangle $ which are not zero and tend to increase in time. Because of this correlation, the model Eq.~\ref{eq:model} filters out any component of the signal (e.g. $\Psi$), which is linearly dependent on $t$ and enters the right side of Eq.~\ref{eq:model}. Therefore, the output position residuals contain only a part of the initial signal amplitude. However, this  does not hamper our statistical analysis, which we performed after the subtraction of $\Psi(t)$ from the observed positions, thus assuming zero input signal and consequently zero output signal. In this way, the impact of correlations is minimized. If the planet does not exist but the subtraction of $\Psi(t)$ was applied, we should detect the inverse signal  $-\Psi(t)$ reduced in amplitude because of the correlation between $\langle \sin \gamma \rangle $ and $\Psi(t)$. This is the case corresponding to the last row of Table~\ref{prob} (Sect.~\ref{sec:discussion}), where the measured signal (expressed by the parameter $Q$) has about half its expected value $-Q$(PS).}

{Fig.~\ref{abcplot} shows the results {in terms of model deviations} in Decl, where the expected signal is largest. The single-frame precision $\sigma_{\rm 1}$ includes errors of photocentre measurements, the reference frame noise, and the atmospheric noise. {It} varies from 0.4 to 0.7~mas depending on seeing. The effect of the vignetted reference field of the 2 Aug epoch (configurations 'a' and 'b') is seen as a degradation of  $\sigma_1$ to over 1~mas. At other epochs, $\sigma_{\rm 1}$ is larger compared to configuration 'c' because of a larger $R_k$.}

\begin{figure}
\begin{center}
\includegraphics*[ width=6.6cm,  height=6.8cm]{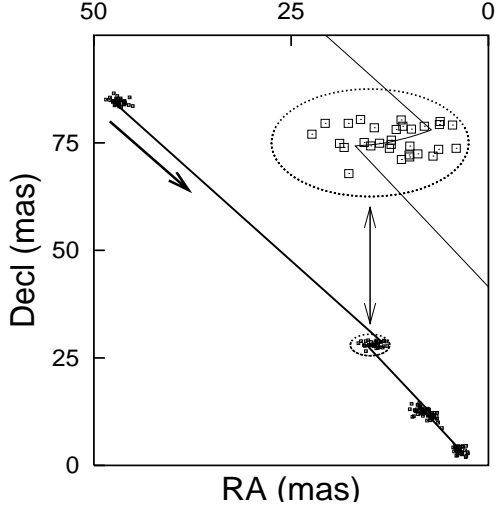} 
\caption{Measured in-frame positions $\hat x-\Phi^{\{ x \}}$, $\hat y-\Phi^{\{ y \}}$ (squares) of VB~10, the model motion (solid line) over the CCD between 8 and 25 Sept (short dataset) defined primarily by proper motion and parallax, and the magnified DCR signature for a single night (right upper corner).}
\label{2D}
\end{center}
\end{figure}

Clear conclusions can be drawn from the short dataset when the model (Eq.~\ref{eq:model2}) is most simple { and does not require incorporation of parallax (Sect.~3.6).}  We considered the cases with the predicted orbital motion subtracted ($\Psi \neq 0$) and with $\Psi = 0$. The epoch average deviations $\langle x \rangle$ and $\langle y \rangle$ are very small and randomly scattered {when} assuming $\Psi = 0$, but display a negative trend in time if $\Psi \neq 0$. Note that small position deviations do not correspond to a 'zero' measurement. Instead, they demonstrate very precise position measurements, which track the proper motion and parallax displacement at the daily rate of 2.7 mas and 5.0 mas in RA and Decl, respectively. The motion of VB~10 over the CCD surface (Fig.~\ref{2D}) for the measurement timespans of 17 days is 46 mas in RA and 86 mas in Decl, {and is dominated by parallax and proper motion}. {DCR effects induce a small-scale scatter in the measured positions of one night} with an amplitude of about 2~mas. {Their structure for a typical night is shown in 5-fold magnification in Fig.~\ref{2D}.}
                  
{Similar computations were performed for the full dataset, where we had to account for the parallax correction $\pi'(d)$ (see Sect.~\ref{sec:col}), thus used the five parameters $x_0$, $y_0$, $\rho$, $d$, and $\pi'$ for fitting of the model Eq.~\ref{eq:model2}. This introduced a strong correlation between $p_x$, $p_y$, and $\Psi$, which reduced the amplitude of the detected signal.}

%%%%%%%%%%%%%%%%%%%%%%%%%
\begin{table}[tbh]
\caption{Epoch deviations $\langle x \rangle$ and $\langle y \rangle$ in RA and Decl of VB~10 for the models with $\Psi=0$ and $\Psi \neq 0$ (Eq.~\ref{eq:model2}).}
\begin{tabular}{@{}r|r@{\quad}r|r@{\quad}r|r@{\quad}r@{\quad}l@{}}
\hline
\hline
            &\multicolumn{2}{c|}{ $\Psi=0$}  &\multicolumn{2}{c|}{ $\Psi \neq 0$} \\% \rule{0pt}{11pt}\\
\hline
MJD         & RA       & Decl        & RA   & Decl    & $\sigma_{\rm N}$ & $\sigma_{\rm sum}$ & $\sigma_{\rm fit}$ \\%\rule{0pt}{11pt}\\ 
55000+      &  (mas)  &  (mas)  &        (mas)  &  (mas)  &  (mas)  &  (mas)  &  (mas)   \\
\hline
        & \multicolumn{6}{c}{ Full dataset, 2 Aug--25 Sept 2009} \\%  \rule{0pt}{11pt}\\
\hline
 45.163 &  -0.26 &   -0.16&       -0.79  &    0.41  &   0.19 &  0.64  &     0.32   \\ %\rule{0pt}{11pt} \\
 82.032 &   0.20 &    0.19&        0.15  &    0.35  &   0.10 &  0.18   &    0.16  \\
 92.993 &  -0.09 &   -0.06&       -0.06  &    0.11  &   0.08 &  0.11   &    0.10 \\
 97.014 &  -0.00 &    0.18&        0.03  &    0.08  &   0.07 &  0.11   &    0.09 \\
 99.011 &  -0.01 &   -0.25&        0.03  &   -0.38  &   0.07 &  0.12   &    0.10 \\
\hline                                
     & \multicolumn{6}{c}{ Short dataset, 8--25 Sept 2009} \\%  \rule{0pt}{11pt}\\
\hline
82.032 &   0.07 &   -0.09  &      0.33 &     0.43 &       0.11&    0.19 &  0.16 \\% \rule{0pt}{11pt}   \\
92.993 &  -0.02 &   -0.04  &     -0.12 &    -0.09 &       0.09&    0.09 &  0.06   \\ 
97.014 &   0.03 &    0.07  &      0.02 &    -0.00 &       0.07&    0.08 &  0.05   \\ 
99.011 &  -0.06 &   -0.01  &     -0.09 &    -0.15 &       0.08&    0.09 &  0.07   \\ 
\hline
\end{tabular}                                                             
\label{ocfull}                                                            
\end{table}                                                               
%%%%%%%%%%%%%%%%%%%%%%%%%%%%%%%%%%%%%%%
                                 
\begin{figure}
\begin{center}
%\begin{tabular}{@{}c@{}c@{}}
\begin{tabular}[b]{@{}c@{}}
\includegraphics*[bb = 95 325 493 468,  width=\linewidth, height=2.95cm]{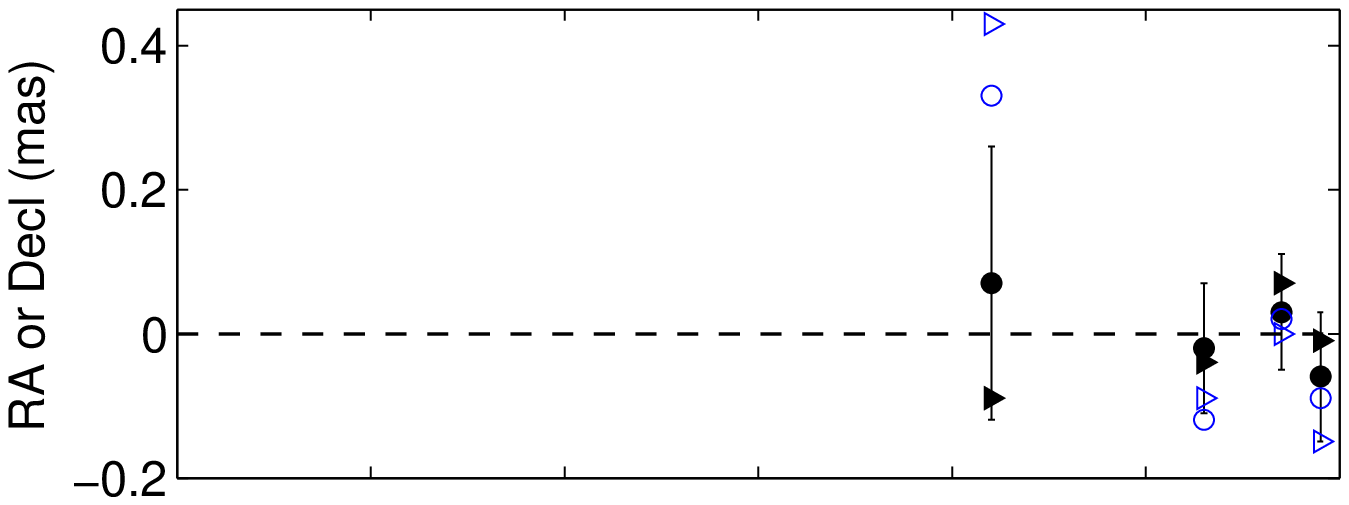}\\
\includegraphics*[bb = 95 306 493 478,  width=\linewidth, height=3.65cm]{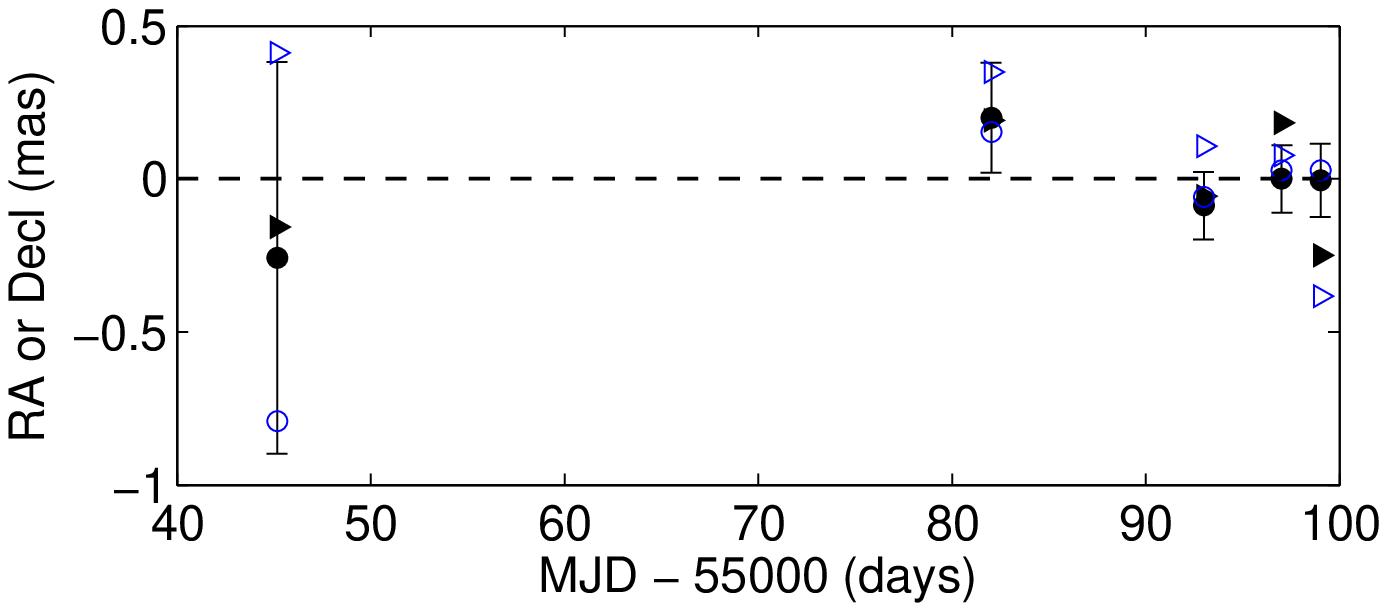} \\
\end{tabular} 
%\includegraphics*[bb = 52 48 186 273, width=2.8cm,  height=6.8cm]{2d.eps} 
%\end{tabular}
\caption{Astrometric residuals in the scenario with planet (open blue symbols) and without planet (filled black symbols) for the short dataset (\emph{top panel}) and the full dataset (\emph{bottom panel}) as a function of time. Circles and triangles mark residuals in RA and Decl, respectively. For clarity, only one errorbar corresponding to $\sigma_{\rm sum}$ is shown at each epoch, but the uncertainties for the four respective measurements are identical.}
\label{plot}
\end{center}
\end{figure}

Table \ref{ocfull} and Fig.~\ref{plot} summarise the results for the epoch residuals $\langle x \rangle$ and $\langle y \rangle$. The astrometric precision is described by a nominal precision $\sigma_{\rm N}$ based on errors in photocentre determination, the reference frame noise and atmospheric noise. $\sigma_{\rm sum}$ also includes error components which dominate at long time spans and originate from the uncertainties in $b$, $\mu'$, pixel scale, and proper motion of VB~10. For the full dataset, it also includes the uncertainty in the parallax of VB~10. $\sigma_{\rm fit}$ is the mathematical expectation of the root-mean-square of $\langle x \rangle$ and $\langle y \rangle$, derived from the least squares fit (Eq.~\ref{eq:model2}).

\section{Discussion}\label{sec:discussion}
Because the signal $\Psi$ was subtracted in Eq.~\ref{eq:model2}, the epoch deviations $\langle x \rangle$ and $\langle y \rangle$ should have {an expectation value of zero and} a random scatter irrespective of the model (with or without planet). {A small dispersion of $\langle x \rangle$ and $\langle y \rangle$ values indicates that the tested hypothesis is correct.} The data shown in Fig.~\ref{plot} and Table~\ref{ocfull} is more consistent with $\Psi = 0$ (the hypothesis of no planetary companion), while incorporating a non-zero planetary signal  $\Psi \neq 0$ increases the scatter of data points, which does not support the existence of {the planet} VB~10b. We estimated the probability $P$ that the planet exists in two ways. 

The first estimate $P( \chi^2)$ was based on the values of $\chi^2=\sum [ \langle x \rangle^2/ \sigma_{\rm sum}^2+ \langle y \rangle^2/ \sigma_{\rm sum}^2]$, where the sum is taken over epochs. The number of degrees of freedom {(D.o.F.)} is equal to the number of epochs minus the number of model parameters {(there are two and three free parameters} for the short and the full dataset, respectively). The probabilities $P( \chi^2)$ corresponding to the $ \chi^2$ values are given in Table~\ref{prob}. For both the short and the full dataset, they are at the equally small{ at the} 2\% level if the model assumes $\Psi \neq 0$. Because the probability of this event is low, the large scatter of $\langle x \rangle$ and $\langle y \rangle$  {is unlikely to be caused by} random noise in the observations.

\begin{table}[tbh]
\caption{Probabilities $P(\chi^2)$, $P(Q)$ and linear trends $Q$ (mas yr$^{-1}$)}
\begin{tabular}{@{}r|r|rcr|rrl@{}}
\hline
\hline
Dataset & $\Psi$    & $\chi^2$ & D.o.F. & $P(\chi^2)$  & $Q_{\alpha}$  & $Q_{\delta}$    & $P(Q)$  \\%   \rule{0pt}{11pt}\\ 
\hline
full     &   0     &10.6     &  7    & 0.16  &   0.67   & -0.35   &  0.36 \\% \rule{0pt}{11pt}\\
full     & $\neq 0$  &18.0    &  7    & 0.02  &   4.14   & -5.38   &  0.02    \\
short   &  0       & 1.9    &  6    & 0.93  &  -2.06   &  2.47   &  0.21  \\
short    & $\neq 0$  & 14.8   &  6    & 0.02&  -7.82   & -11.32  & $5 \cdot 10^{-4}$     \\

\hline
\end{tabular}
\label{prob}
\end{table}

Secondly, we noticed that the observed residuals $\langle x \rangle$ and $\langle y \rangle$ {show a linear trend instead of a} random distribution. This can be caused by random errors in the observations, but also may indicate a wrong value of the signal $\Psi(t)$ subtracted in Eq.~\ref{eq:model2}. {In the latter case, we would expect that the residuals $\langle x \rangle$ and $\langle y \rangle$ show a linear dependence $-Qt$ opposite to $\Psi(t)$ (see Eq.~\ref{eq:appr}).}

A large negative trend is seen for the short dataset residuals computed with $\Psi \neq 0$ (Fig.~\ref{plot}), whereas they {should be near} zero to support the planetary hypothesis. In contrast, a much smaller trend is observed when assuming $\Psi=0$ (no planet).  {For instance, the measured values of $Q$  given in Table~\ref{prob} for the short dataset and $\Psi \neq 0$ are -7.82 and -11.32 mas yr$^{-1}$ in RA and Decl, respectively. If the null hypothesis $\Psi \neq 0$ is correct, their expected values are 0 and 0 mas yr$^{-1}$, and -13.4 and -22.4 mas yr$^{-1}$  if the hypothesis is wrong ($\Psi = 0$).} In the last case, the measured $Q$ values are always smaller than the values $Q({\rm PS})$ induced by the orbital motion, because their magnitude is damped by correlations between  $\Psi(t)$ and model parameters (cf. Sect.~\ref{sec:posres}).

Because the $\chi^2$ criterion is not the most efficient one for the characterisation of linear signals, we developed another, more powerful approach to {obtain} an alternative estimate of $P$. We considered $Q$ as quantities describing the dispersion of $\langle x \rangle$ and $\langle y \rangle$ better than $\chi^2$ and performed Monte Carlo simulations to see if the observed features can{ be explained by }random noise in the observations. We simulated each data frame by a random Gaussian noise with root-mean-square of $\sigma_{\rm 1}$ and added components modelling the errors in $\mu'$, $b$, and in the proper motion and parallax of VB~10.
{In addition, we included a 1 mas yr$^{-1}$ error to account for the uncertainties in the zero point of proper motion (cf. Sect.~\ref{sec:refframes}).}

{After fitting the data with the model Eq.~\ref{eq:model2}, we searched the residuals for linear trends in time and estimated their coefficients $Q^*$. }The false-alarm probabilities $P(Q)$ that a random noise produces
$Q^*$ values larger than the observed ones are given in Table~\ref{prob}. In the case of the short dataset and $\Psi \neq 0$, we find that the observed linear trends can be explained by Gaussian noise with a probability of $P(Q)= 5 \cdot 10^{-4}$, which does not support the existence of the planet. For the full dataset, we find $P(Q)=0.02$, which is not sufficient to draw definite conclusions. In spite of the longer timespan and therefore seemingly better conditions to characterise a planetary signal, the full dataset does not provide a better contraint {because of} the uncertainty in $\mu'$, whose contribution increases with time. Besides, the planetary signal is substantially damped because of its strong correlation with the colour correction $\pi'$ used in Eq.~(\ref{eq:model2}) as a free model parameter.

\section{Conclusions}
We conclude that the presence of {the announced planet around VB~10 }is not supported by astrometry. Even  assuming half the  planetary mass (i.e. 3.2 M$_{\rm J}$), simulations give  a low  false alarm probability $P(Q)= 0.023$, which does still not support the existence of VB~10b. Our result obtained from astrometry alone is in agreement with the conclusion of \cite{Bean} and \cite{Anglada} based on RV data.

{This study is the first application of the FORS2 camera} for the search of exoplanets by means of optical astrometry.{ Because of the} high astrometric precision of FORS2, the availability of external STEPS-based proper motion and parallax of VB~10, and the large expected orbital signal, {it was possible to perform the verification of the planetary companion hypothesis within the extremely short observation period of 17 days}, which is unusual for astrometric works of this type. The successful use of the STEPS data for the reduction of FORS2 observations is justified only { because we verified that the STEPS and FORS2 proper motion reference frames are consistent within 1 mas~yr$^{-1}$ uncertainty.}

We have demonstrated a mean nominal precision of 0.09 mas per epoch of FORS2/VLT observation for data of reasonable quality, despite problems caused by the uncertainty {in the} LADC position. This precision {is} sufficient for astrometric detection of planets around ultracool dwarfs.

\begin{acknowledgements}
We thank Dr. G. Anglada-Escud\'e whose comments have helped to improve the paper. PF and NCS would like to thank the support by the European Research Council/European Community under the FP7 through a Starting Grant, as well as the support from Funda\c{c}\~ao para a Ci\^encia e a Tecnologia (FCT), Portugal, in the form of a grant with reference PTDC/CTE-AST/098528/2008. NCS would further like to thank the support from Funda\c{c}\~ao para a Ci\^encia e a Tecnologia (FCT), Portugal, through a Ci\^encia\,2007 contractfunded by FCT/MCTES (Portugal) and POPH/FSE (EC).
\end{acknowledgements}
\bibliographystyle{aa}
\bibliography{vb10bib}

\end{document}